\documentclass[11pt]{article}
\title{Exact uncertainty relations}
\author{Michael J. W. Hall\\
Theoretical Physics, IAS\\Australian National University\\
Canberra ACT 0200, Australia}
\date{}
\begin{document}
\maketitle
%\newpage

\begin{abstract}
The Heisenberg inequality $\Delta X\Delta P\geq\hbar/2$ 
can be replaced by an exact {\it equality}, for suitably
chosen measures of position and momentum uncertainty, which is valid for
{\it all}
 wavefunctions.  The 
statistics of complementary observables are thus connected by an
``exact" uncertainty relation. 
Results may be generalised to angular momentum and phase, photon
number and phase, time and
frequency, and to states described by density operators.  Connections to
energy bounds, entanglement, Wigner functions, and 
optimal estimation of an observable from the measurement of a second
observable are also given. 
\end{abstract}
\renewcommand{\thesection}{\Roman{section}}
\renewcommand{\thesubsection}{\Alph{subsection}}
\newpage
\section{INTRODUCTION}
One of the most striking features of quantum mechanics is the property that
certain observables cannot simultaneously be assigned arbitrarily precise
values.  This property does not compromise claims of completeness for the
theory, since it may consistently be asserted that such observables cannot
simultaneously be {\it measured} to an arbitrary accuracy \cite{heisbohr}.
The Heisenberg inequality
\begin{equation} \label{heis}
\Delta X \Delta P \geq \hbar /2
\end{equation}
is therefore generally taken to 
reflect an essential incompleteness in the applicability of classical
concepts of position and momentum to physical reality.

It was recently noted that this fundamental inequality can 
be greatly strengthened - the degree to which classical concepts are
inapplicable can, surprisingly,
be quantified exactly.  In particular, one may
define a measure of position uncertainty $\delta X$ (which arises naturally in
classical statistical estimation theory), and a measure of nonclassical
momentum uncertainty $\Delta P_{nc}$ (which arises from a natural
decomposition of the momentum operator), such that \cite{hall} 
\begin{equation} \label{ex}
\delta X \Delta P_{nc} = \hbar /2 
\end{equation}
for {\it all} wavefunctions.
Such an equality may be regarded as an {\it exact} uncertainty relation, and
may be shown to imply the usual Heisenberg inequality Eq. (\ref{heis}).
Thus, perhaps paradoxically, the uncertainty principle
of quantum mechanics may be given a precise form.

In Ref. \cite{hall} 
the above exact uncertainty relation was merely noted
in passing, with the emphasis being on
other properties of $\delta X$ and
$\Delta P_{nc}$. 
Similarly, while the very existence of
an exact form of the uncertainty principle was recently shown
to provide a sufficient basis for 
moving from classical equations of motion to the Schr\"{o}dinger equation
\cite{hallreg}, the corresponding exact uncertainty relation Eq. (\ref{ex})
was only briefly mentioned.  The purpose of this paper, therefore, is to study
the physical significance of Eq. (\ref{ex}) in some detail, 
including its extensions to 
other pairs of conjugate observables and to general states described by
density operators.

In the following section it is shown that quantum observables such as
momentum, position, and photon number have a natural decomposition, into
the sum of a classical and a 
nonclassical component.  The classical component 
corresponds to the best possible measurement of the
observable, on a given state, which is compatible with measurement of
the {\it conjugate} observable.  Complementarity implies
that the classical component cannot be equivalent to
the observable itself, i.e., there is in general an nontrivial
{\it non}classical component.  It is this nonclassical component which
reflects the mutual incompatibility of pairs of conjugate observables,
and the magnitude of which appears in the exact uncertainty relations to
be derived [e.g., $\Delta P_{nc}$ in Eq. (\ref{ex})]. 
The decomposition into classical and nonclassical components is also related
in a natural manner to quantum continuity equations and to
quasiclassical properties of the Wigner function.

In Sec. III a measure of uncertainty is defined for continuous
random variables such as position, which plays a fundamental role in classical 
estimation theory and in Gaussian diffusion processes.  This 
measure, the ``Fisher length'' of the variable, may of course be
calculated for quantum observables as well, and appears as $\delta X$ in
the exact uncertainty relation in Eq. (\ref{ex}).

The ingredients of classical/nonclassical decompositions and
Fisher lengths are combined in Sec. IV to obtain a number of exact uncertainty
relations, such as Eq. (\ref{ex}) and the equality
\[ \delta \Phi \Delta N_{nc}=1/2\]
for phase and photon number, valid for all pure states.  These relations
generalise to {\it in}equalities for states described by density operators,
and are far stronger than the corresponding Heisenberg-type
inequalities.  A simple proof is given of the property that localised
quantum states have infinte kinetic energy (arising from the 
contribution of the nonclassical momentum component), and
it is shown that a bound on Fisher length leads to an
entropic lower bound for the groundstate energies of quantum systems.

In Sec. V it is shown that the decomposition of an observable of
a given quantum system into
classical and nonclassical components is essentially nonlocal in nature,
being dependent in general on manipulations performed on a  
second system with which the first is entangled.  The significance of
the relevant exact uncertainty relations is discussed, with particular
reference to EPR-type states.

A formal generalisation of exact uncertainty relations, to
arbitrary pairs of quantum observables, is noted in Sec. VI.  Moreover,
it is shown that a result of Ivanovic \cite{ivan}, for complete sets of
mutually complementary observables on finite Hilbert spaces (such as the
Pauli spin matrices), may be reinterpreted as an exact uncertainty
relation for the ``collision lengths'' of the observables.

Conclusions are given in Sec. VII.
\newpage
\section{CLASSICAL AND NONCLASSICAL COMPONENTS OF QUANTUM OBSERVABLES}

\subsection{Classical momentum}

The nonclassical momentum uncertainty 
$\Delta P_{nc}$ appearing in Eq. (\ref{ex}) is
defined via a natural decomposition of the momentum observable $P$ into
``classical'' and ``nonclassical'' components,
\begin{equation} \label{pdecomp}
P = P_{cl} + P_{nc} .
\end{equation}
This decomposition is state-dependent, and will be defined explicitly
further below.  In particular, it will be shown that {\it the classical
component, $P_{cl}$, corresponds to the best possible 
estimate of momentum, for a given
quantum state, which is compatible with a position measurement}. 
It will be seen further below that $P_{cl}$ is also related to
the momentum flow in the classical continuity equation for the
position probability density, and to an average momentum arising naturally
from quasiclassical properties of the Wigner function.  However, it is
the ``best estimate'' interpretation above that provides the most
general basis for generalisation to other observables. 

As a starting point, recall that in classical mechanics one can
simultaneously obtain precise values for position and momentum, whereas
in quantum mechanics one must choose to accurately measure either one or
the other.  It is therefore reasonable to ask the following question:
If I measure one of these observables precisely, on a known quantum
state, then what is the best
estimate I can make for the value of the other observable? Such an
estimate of momentum from the measurement of position will be called a
{\it classical} estimate of $P$, since it assigns simultaneous values to
$X$ and $P$.

It will be shown that the {\it best} classical estimate of $P$, given
the measurement result $X=x$ on a quantum system described by wavefunction
$\psi (x)$, is given by
\begin{equation}
\label{pclpsi}
P_{cl}(x) = \frac{\hbar}{2i}
\left( \frac{\psi^\prime (x)}{\psi(x)}-\frac{\psi^{*\prime} (x)}{\psi^*(x)}
\right) = \hbar [\arg \psi (x)]^\prime .
\end{equation}
More generally, for a quantum system described by density operator
$\rho$, one has
\begin{equation} \label{pcl}
P_{cl}(x) := \frac{\langle x|P\rho +\rho P|x\rangle/2}{\langle x|\rho
|x\rangle} , 
\end{equation}
which reduces to the first expression for $\rho =|\psi\rangle\langle\psi
|$.  

The experimentalist's procedure for measuring the classical momentum
component for state $\rho$ is thus to (i) prepare the system in
state $\rho$; (ii) measure the position $X$; and (iii) for result $X=x$
calculate $P_{cl}(x)$. 
Note that this is equivalent to measurement of the
Hermitian operator 
\begin{equation} \label{pclop}
P_{cl} = \int dx\,  P_{cl}(x)|x\rangle\langle x| 
\end{equation}
on state $\rho$, which by construction commutes with $X$. Hence the
statistics of $P_{cl}$ are determined by those of $X$. 
As stated above, this procedure yields the best
possible estimate of the momentum of the system that is compatible with
simultaneous knowledge of the position of the system.

To prove that $P_{cl}(x)$ provides the best classical estimate of
$P$, consider some general classical estimate
for momentum for
state $\rho$.  Since this estimate must by definition be compatible
with the measurement of $X$, it is formally equivalent to the measurement of 
some operator $\tilde{P} = \int dx\,\tilde{P}(x)|x\rangle\langle x|$.
The average error of the estimate may therefore be quantified by
the mean deviation of $\tilde{P}$ from the momentum operator $P$,
\begin{equation} \label{error1}
{\cal E}_P := \langle (P-\tilde{P})^2\rangle ,
\end{equation} 
where $\langle A\rangle$ denotes the expectation value ${\rm tr}[\rho A]$.  

Using the cyclic property of the trace operation
and evaluating the trace in the position representation gives
\begin{eqnarray*}
\langle \tilde{P}P+P\tilde{P}\rangle & = & \int dx\,\langle x|\tilde{P}
P\rho + \rho P\tilde{P}|x\rangle\\
& = & \int dx\, \tilde{P}(x)\langle x|P\rho +\rho P|x\rangle\\
& = & 2\int dx\, \langle x|\rho |x\rangle \tilde{P}(x) P_{cl}(x) =
2\langle\tilde{P}P_{cl}\rangle ,
\end{eqnarray*}
and hence
\begin{eqnarray}
{\cal E}_P & = & \langle P^2\rangle +\langle \tilde{P}^2\rangle -
2\langle\tilde{P}P_{cl}\rangle\nonumber \\ \label{error2}
& = & \langle P^2\rangle - \langle P_{cl}^2\rangle +\langle (\tilde{P}
- P_{cl})^2\rangle \label{error} .
\end{eqnarray}
Since the last term is positive, the average error is 
minimised by the choice $\tilde{P}=P_{cl}$ as claimed.

\subsection{Nonclassical momentum}

The nonclassical momentum component $P_{nc}$ is now defined via
Eq. (\ref{pdecomp}), as the difference of the quantum
momentum $P$ and the classical momentum $P_{cl}$.  
From Eq.~(\ref{pcl}) one finds that the expectation values of
the observables $P$ and $P_{cl}$ are always equal (for the corresponding
state $\rho$), and so
\begin{equation} \label{pav}
\langle P\rangle = \langle P_{cl}\rangle ,\hspace{1cm}\langle
P_{nc}\rangle = 0 . 
\end{equation}
The quantum momentum $P$ in Eq. (\ref{pdecomp}) 
can therefore also be interpreted as the sum of
an average momentum $P_{cl}$, 
and a nonclassical momentum fluctuation $P_{nc}$.

The magnitude of this nonclassical fluctuation is simply related to the
minimum average error for a classical estimate:  choosing $\tilde{P}=P_{cl}$ 
in Eqs. (\ref{error1}) and (\ref{error2}) yields 
\begin{equation} \label{pdiff}
\langle P_{nc}^2\rangle = {\cal E}_P^{\rm min}
= \langle P^2\rangle - \langle P_{cl}^2\rangle .
\end{equation} 
It will be seen that, as a consequence of the exact uncertainty relation
Eq. (\ref{ex}), this minimum error does not vanish for any state (although it
may be arbitrarily small), and hence there is always a residual amount
of nonclassicality.
Note from Eqs. (\ref{pav}) and (\ref{pdiff}) that the 
fluctuation strength $\Delta P_{nc}$ in Eq. (\ref{ex}) is a fully operational
quantity, as it may be determined from the measured distributions of $P$
and $P_{cl}$ (and hence from the measured distributions of $P$ and $X$). 

Finally, since the decomposition into
classical and nonclassical components is state dependent,  
$P_{cl}$ and $P_{nc}$ should, strictly
speaking, explicitly indicate their dependence on a given state $\rho$
(e.g., via the notation $P_{cl}^\rho$  and $P_{nc}^\rho$ respectively).
This would in particular be necessary if one wished to evaluate 
expectation values such as ${\rm tr}[\sigma P_{cl}^\rho]$, for some density
operator
$\sigma$ other than $\rho$.
However, expectation values will in fact only be
evaluated for the corresponding state $\rho$ throughout this paper,
and hence explicit notational dependence on the state may be conveniently
dispensed with, without
leading to ambiguity.

\subsection{Physical significance}

It is seen that the classical momentum is the closest possible observable 
momentum observable $P$ (in a statistical sense), 
under the constraint of being co-measurable with the conjugate
position observable $X$. The nonclassical momentum is then simply
defined as the difference between the quantum momentum and the classical
momentum.  A similar approach can be used to define corresponding decompositions
of the position, angular momentum, and photon number observables.

The decomposition in Eq. (\ref{pdecomp}) attempts to demarcate
classical and nonclassical
momentum properties.
It is therefore reasonable to hope that the {\it nonclassical}
component $P_{nc}$ in particular might play a fundamental role in
describing the
essence of what is ``quantum'' about quantum mechanics.  This is indeed
the case.  A derivation of the Schr\"{o}dinger equation
as a consequence of adding a nonclassical momentum fluctuation to a
classical
ensemble (with strength inversely proportional to the uncertainty in
position),
has recently been given
\cite{hallreg}.  In this paper it will be shown that the nonclassical
components of quantum observables 
satisfy exact uncertainty relations,
such as Eq. (\ref{ex}), and hence allow one to {\it precisely} quantify
the fundamental uncertainty principle of quantum mechanics.  
It will further be shown that the
decomposition of observables into classical and nonclassical components
helps to distinguish between local and nonlocal features of quantum
entanglement.

Several formal properties further support the physical significance
of the decomposition in Eq. (\ref{pdecomp}).  
First, the classical and
nonclassical components are linearly uncorrelated, i.e.,
\begin{equation} \label{pvar}
{\rm Var} P = {\rm Var} P_{cl} + {\rm Var} P_{nc} , 
\end{equation}
as follows immediately from Eqs. (\ref{pav}) and
(\ref{pdiff}).  This
implies a degree of statistical, and hence physical,
independence for $P_{cl}$ and $P_{nc}$.  The same
equations imply that the kinetic energy splits into a
classical part, $\langle P_{cl}^2/(2m)\rangle$ and a nonclassical 
part, $\langle P_{nc}^2/(2m)\rangle$. Note from Eq. (\ref{pclpsi}) that 
the former contribution vanishes for stationary states, 
leaving only a nonclassical
contribution to the kinetic energy of such states. 

Second, the classical momentum
component commutes with the conjugate
observable $X$ while the nonclassical component does not, i.e.,
\begin{equation}
[X,P_{cl}]=0,\hspace{1cm} [X,P_{nc}] = i\hbar .\nonumber 
\end{equation}
Hence it is the {\it nonclassical} component of $P$ which generates
the fundamental quantum property $[X,P]=i\hbar$. 

Third, when the decomposition is
generalised to more than one dimension (see Sec. V.A), one finds 
for pure states that
the commutativity property $[P^j,P^k]=0$ for the vector components of
momentum is preserved by the decomposition, i.e.,
\begin{equation}\label{commcomp}
[P_{cl}^j, P_{cl}^k] = 0 = [P_{nc}^j, P_{nc}^k].\nonumber
\end{equation}

Fourth, the classical momentum $P_{cl}(x)$ associated with position $X=x$
appears in the continuity equation \cite{merz} 
\[
\partial |\psi |^2/\partial t + (\partial/\partial x)\left[ |\psi |^2
m^{-1} P_{cl}(x)\right] = 0
\]
following from the Schr\"{o}dinger equation, 
and hence $P_{cl}$ corresponds to the flow
momentum of a classical ensemble of particles described by probability
density $|\psi |^2$. 
This property suggests an alternative ``dynamical'' approach
to defining classical/nonclassical decompositions such as Eq.
(\ref{pdecomp}).  However, such an approach can generally only
be applied
to systems with Hamiltonians that are quadratic in the observable to be
decomposed.
 
Finally a ``quasiclassical'' approach to the decomposition in Eq.
(\ref{pdecomp}) is noted, based on an analogy between
classical phase space distributions and the Wigner function \cite{wig}
\begin{equation} \label{wig}
W(x,p) := (2\pi\hbar)^{-1}\int d\xi 
e^{-ip\xi/\hbar}\langle x-\xi/2|\rho |x+\xi/2\rangle ,
\end{equation}
where the latter 
behaves, at least to some extent, 
like a joint probability density for position and momentum \cite{wig}. 

Now, for any true classical joint probability density $w(x,p)$ on phase
space, the {\it average} classical 
momentum associated with position
$x$ is given by 
$p_{cl}(x)=\int dp\, p\, {\rm prob}(p|x)$, 
where ${\rm prob}(p|x)$ denotes
the conditional probability that the momentum is equal to $p$ at
position $x$, i.e.,
${\rm prob}(p|x) = w(x,p)/\int dp\, w(x,p)$. 
The average classical momentum at position $x$ is thus 
\[
p_{cl}(x) = \frac{\int dp\, p w(x,p)}{\int dp \, w(x,p)}.
\]

For {\it quantum} systems this immediately suggests defining an analogous 
average classical 
momentum associated with position $x$, via replacement of $w(x,p)$ by the
Wigner function \cite{takbrown}, to give
\begin{equation} \label{qav}
P_{cl}(x) := \frac{\int dp\, p W(x,p)}{\int dp \, W(x,p)}. 
\end{equation}
Remarkably, as shown elsewhere \cite{eurlong},
this is in fact {\it equivalent} to the definition in Eq.
(\ref{pcl}) !  Note that this quasiclassical approach reinforces the 
interpretation of Eq. (\ref{pav}), that the momentum of a
quantum particle comprises a nonclassical fluctuation $P_{nc}$ about a classical
average $P_{cl}$.

One may similarly define a corresponding classical component for
the position observable $X$,
by interchanging the roles of $x$ and $p$ in Eq. (\ref{qav}). 
This agrees with the analogous definition based on Eq. (\ref{pcl}),
corresponding to the more generally applicable ``best estimate'' approach, 
and also with the definition given 
in Ref. \cite{hall} based on a semiclassical continuity equation.
\newpage
\section{FISHER LENGTH}

The uncertainty measure $\Delta P_{nc}$ in Eq. (\ref{ex}) is now well
defined - it is the rms uncertainty of the nonclassical momentum component
$P_{nc}$.  However, it still remains to define the measure of position
uncertainty $\delta X$ in Eq. (\ref{ex}).  This is done below for the
general case of observables taking values over the entire set
of real numbers, such as position and momentum. 
Note that $\delta X$ is a purely {\it
classical} measure of uncertainty, requiring no reference to quantum
theory whatsoever.  

For a random variable $X$ which takes values over the whole range of
real numbers, there are of course many possible ways to quantify the
spread of the corresponding distribution $p(x)$.  Thus, for example, one
may choose the rms uncertainty $\Delta X$, the collision length $1/\int
dx \, p(x)^2$ \cite{hell}, or the ensemble length $\exp [-\int dx\, p(x) \ln
p(x)]$ \cite{hallvol}.  All of these examples have the desirable
properties of having the same units as $X$, scaling linearly with $X$, being
invariant under translations of $X$, and
vanishing in the limit as $p(x)$ approaches a delta function.

A further uncertainty measure satisfying the above
properties is
\begin{equation} \label{fl}
\delta X := \left[ \int_{-\infty}^{\infty} dx \, p(x) \left( \frac{d\ln
p(x)}{dx}\right)^2\right]^{-1/2}  .
\end{equation}
While this measure may appear unfamiliar to physicists, it is in fact
closely related to the well known Cramer-Rao inequality that lies at
the heart of statistical estimation theory \cite{cox}: 
\begin{equation} \label{cramer}
\Delta X \geq \delta X . 
\end{equation}
Thus $\delta X$ provides a lower bound for $\Delta X$. 
Indeed, more generally, $\delta X$ provides 
the fundamental lower bound for the rms
uncertainty of {\it any} unbiased estimator for $X$ \cite{cox}.  The bound 
in Eq. (\ref{cramer}) is
tight, being saturated if and only if $p(x)$ is a Gaussian distribution.

Eq. (\ref{cramer}) is more usually written in the form ${\rm Var} X\geq1/F_X$,
where $F_X=(\delta X)^{-2}$ is the ``Fisher information'' associated
with translations of 
$X$ \cite{cox,fish,stam, dembo}.  It is hence appropriate to refer to $\delta
X$ as the {\it Fisher length}.  
From Eq. (\ref{fl}) it is seen that the Fisher length may be regarded
as a measure of the length scale over which $p(x)$ [or, more precisely,
$\ln p(x)$] varies rapidly.

Basic properties of the Fisher length are: (i) $\delta Y=\lambda\delta X$ for
$Y=\lambda X$; (ii) $\delta X\rightarrow 0$ as $p(x)$ approaches a delta
function; (iii) $\delta X \leq \Delta X$ with equality only for Gaussian
distributions; and (iv) $\delta X$ is finite for all distributions.  
This last property follows since the integral in Eq. (\ref{fl}) can
vanish only if $p(x)$ is constant everywhere, which is inconsistent with
$\int dx\, p(x)=1$.

The Fisher length has the unusual feature that it depends on the
derivative of the distribution.   Moreover, for this reason it vanishes for
distributions which are discontinuous - to be expected from the above 
interpretation of $\delta X$, since such distributions vary
{\it infinitely} rapidly over a {\it zero} 
length scale ($\delta X=0$ may be shown by replacing such
a discontinuity at point $x_0$ by a linear interpolation over an
interval $[x_0-\epsilon, x_0+\epsilon]$ and taking the limit
$\epsilon\rightarrow 0$).  The Fisher length also vanishes
for a distribution that is zero over some interval (since $\ln p(x)$ in
Eq. (\ref{fl}) changes from $-\infty$ to a finite value over any
neighbourhood containing an endpoint of the interval).  
While these features imply that
$\delta X$ is not a particularly useful uncertainty measure for such
distributions (similarly, $\Delta X$ is not a particularly useful measure
for the Cauchy-Lorentz distribution $(a/\pi)(a^2+x^2)^{-1}$), they
are {\it precisely} the features that lead to a simple proof that the
momentum uncertainty is infinite for any quantum system with a position
distribution that is discontinuous or vanishes over some interval (as will
be shown in Sec. IV).

One further property of Fisher length worthy of note is its alternative
interpretation as a ``robustness length''.  In particular, suppose that
a variable described by $p(x)$  is subjected to a Gaussian diffusion
process, i.e., $\dot{p}=\gamma p''+\sigma p'$ for diffusion constant $\gamma$ 
and drift velocity $\sigma$.  
It then follows from Eq. (\ref{fl}) and de Bruijn's
identity \cite{stam} that the rate of entropy increase is given by
\begin{equation} 
\dot{S} = \gamma /(\delta X)^2 .
\end{equation}
Since a high rate of entropy increase corresponds to a rapid spreading of the
distribution, and hence nonrobustness to diffusion, this inverse-square law
implies that the Fisher length $\delta X$ is a direct measure of
robustness.  Hence $\delta X$ may also be referred to as a
{\it robustness length}.  This characterisation of robustness is 
explored for quantum systems in Ref. \cite{hall}.

Finally, note that Fisher length is not restricted to position
observables, but may be calculated as per Eq. (\ref{fl}) for any
observable which takes values over the entire set of real numbers, such
as momentum.  A Fisher length having similar properties may also be
defined for periodic observables such as phase \cite{eurlong}. 
\newpage
\section{EXACT UNCERTAINTY RELATIONS}

\subsection{Position and momentum}

In the previous two sections the quantities $\Delta P_{nc}$ and $\delta X$
have been motivated and discussed on completely independent grounds.
One is a measure of uncertainty for the nonclassical component
of momentum, while the other is a measure of uncertainty for position that
appears naturally in the contexts of classical statistical estimation
theory and Gaussian diffusion processes.

It is a remarkable fact that for all pure states these two quantities are
related by the simple equality
in Eq. (\ref{ex}), repeated here for convenience:
\begin{equation} \label{rex}
\delta X \Delta P_{nc} = \hbar/2 .
\end{equation}
Thus the Fisher length of position is inversely proportional to the
strength of the nonclassical momentum fluctuation.  
Noting from Eqs. (\ref{pvar})
and (\ref{cramer}) that $\Delta P\geq\Delta P_{nc}$ and $\Delta X\geq
\delta X$ respectively, the Heisenberg uncertainty relation
\begin{equation} \label{rheis}
\Delta X \Delta P \geq \hbar /2
\end{equation}
is an immediate consequence of this {\it exact}
quantum uncertainty relation.

The existence of an exact uncertainty relation for position and momentum
statistics greatly strengthens the usual statement of the uncertainty principle,
from inequality to equality, and hence the measures of uncertainty in
Eq. (\ref{rex}) may regarded as more fundamental in nature than those in
Eq. (\ref{rheis}).  Moreover, the phase space area $\hbar/2$ is promoted in
status, from
a mere lower bound on joint uncertainty to an invariant quantity which
precisely
characterises the joint uncertainty of {\it every} wavefunction.

Thus, consider an ensemble of systems
described by state $\psi$, on which independent measurements of $X$ and
$P$ are made (on different subensembles).  From these measurements
one can determine the statistics of $X$ and $P$ [and hence also the
statistics of $P_{cl}$ and the
variance of $P_{nc}$, from Eqs. (\ref{pclop}) and (\ref{pvar})
respectively]. 
The Heisenberg uncertainty relation connects these statistics via an
inequality - if one calculates $\Delta P$, then one knows 
only
that $\Delta X \geq \hbar /(2\Delta P)$, where the difference between
the lefthand and righthand sides depends on the particular wavefunction
$\psi$ describing the ensemble.  In contrast, the exact uncertainty
relation provides an invariant equality connecting the statistics, 
where if one calculates the nonclassical momentum fluctuation $\Delta
P_{nc}$, then one knows immediately that the Fisher length $\delta X$ is 
{\it precisely} equal to $\hbar/(2\Delta P_{nc})$, regardless of the particular
wavefunction.

A simple proof of Eq. (\ref{rex}) was given in Ref. \cite{hall}; a more
general result, valid for density operators, is proved below.
Before proceeding to the proof, however, several simple consequences 
of the exact uncertainty
relation in Eq. (\ref{rex}) will be noted. 

First, recalling that $\delta X$ vanishes for position distributions
that are discontinuous or are zero over some interval (see Sec. III),
it follows immediately from Eq. (\ref{rex}) that $\Delta P_{nc}$ is
infinite in such cases.  From Eq. (\ref{pvar}) the momentum uncertainty $\Delta
P$ is then also infinite.  Note that this conclusion {\it
cannot} be derived from the Heisenberg inequality Eq. (\ref{rheis}), nor
from the entropic uncertainty relation for position and momentum
\cite{bbm}. The exact uncertainty relation Eq. (\ref{rex}) is thus
significantly stronger than the latter inequalities.  

A second related consequence worth mentioning is a simple proof that 
any well-localized pure state, i.e., one for
which the position distribution vanishes outside some finite interval,
has an infinite energy (at least for any potential energy that is bounded
below at infinity). This is immediately implied by the property 
\begin{equation}\label{energy}
E = (8m)^{-1}\hbar^2(\delta X)^{-2}
+ \langle P_{cl}^2\rangle /(2m) + \langle V(x)\rangle
\end{equation}
(following from Eqs. (\ref{pdiff}) and (\ref{rex})), 
and noting that $\delta X=0$ for such states.
Note that this ``paradox'' of standard quantum mechanics 
(that there are no
states which are both well-localised and have finite energy) is
a consequence of the simple external potential model, rather than of some deep
incompleteness of the theory.  Note also that this property is purely
quantum in nature, since the divergent term - the nonclassical part of
the kinetic energy - vanishes in the limit
$\hbar\rightarrow 0$.

Third, the property $\delta X<\infty$ (see Sec.
III) immediately implies from the exact uncertainty relation Eq.
(\ref{rex}) that $\Delta P_{nc}$ can never vanish, i.e.,
\begin{equation}\label{posit}
\Delta P_{nc} > 0 .
\end{equation}
Thus all pure states
necessarily have a nonzero degree of nonclassicality associated with
them \cite{footconj}.  This result is intuitively appealing, and
provides further support for
the physical significance of the classical and nonclassical 
components.

Fourth, for all {\it real} wavefunctions $\psi (x)$, including energy
eigenstates, one has $P_{cl}\equiv 0$ from Eq. (\ref{pclpsi}). Hence 
the exact uncertainty relation reduces to the simpler identity 
\begin{equation} \label{real}
\delta X \Delta P = \hbar /2 .
\end{equation}
This result holds more generally whenever the phase of $\psi$ is at most
linear in $x$.

Eq. (\ref{rex}) for pure states will now be proved as a special case of
the more general {\it inequality}
\begin{equation} \label{rhoeur}
\delta X\Delta P_{nc}\geq \hbar/2,
\end{equation}
holding for states described by density operators.  
While not an exact uncertainty relation, this inequality is
still much stronger than the corresponding Heisenberg inequality in Eq.
(\ref{heis}).  Not only is it saturated for {\it all} pure states (not
just the ``minimum uncertainty'' states), but it implies that 
generalisations of the above consequences hold for {\it any} quantum state.

Inequality (\ref{rhoeur}) is an immediate consequence of 
Eq. (\ref{pdiff}) and the relations
\begin{equation}\label{rhoeq}
\frac{\hbar^2}{4(\delta X)^2} + \langle P_{cl}^2\rangle = \int dx\,
\frac{|\langle x|P\rho |x\rangle|^2}{\langle x|\rho |x\rangle}
\leq \langle P^2\rangle  , 
\end{equation}
which hold for all density operators $\rho$.  The equality in Eq.
(\ref{rhoeq}) is obtained by substituting Eqs. (\ref{pcl})
and (\ref{pclop}) for the classical momentum component $P_{cl}$,
and the representation
\begin{equation} \label{fisho}
(\delta X)^{-2} = -\frac{1}{\hbar^2}\int dx \, \frac{\langle
x|P\rho-\rho P|x\rangle^2}{\langle x|\rho |x\rangle} ,
\end{equation}
for the Fisher length, following from the definition of $\delta X$ in 
Eq. (\ref{fl}) and the
identity $(d/dx)\langle x|A|x\rangle=(i/\hbar)\langle x|[P,A]|x\rangle$
(derived by expanding in momentum eigenkets). 
The {\it in}equality in Eq. (\ref{rhoeq}) is obtained by defining the
states $|\mu\rangle =\rho^{1/2}P|x\rangle$,
$|\nu\rangle=\rho^{1/2}|x\rangle$, and using the Schwarz inequality 
\[
|\langle x|P\rho |x\rangle |^2 = |\langle\mu |\nu\rangle |^2 \leq
\langle\mu |\mu\rangle\langle\nu |\nu\rangle = \langle x|P\rho
P|x\rangle\langle x|\rho |x\rangle . \]
Remarkably, for the special case of a pure state, direct substitution of 
$\rho=|\psi\rangle\langle\psi |$ into the integral in Eq. (\ref{rhoeq})
yields equality on the righthand side, and hence the exact uncertainty 
relation Eq. (\ref{rex}).

Finally, note that whereas the Heisenberg inquality Eq.
(\ref{rheis}) is symmetric with respect to position and momentum, this
symmetry is broken by the exact uncertainty relation Eq. (\ref{rex}).
Instead, one has {\it two} (symmetrically related) exact uncertainty
relations, given by Eq. (\ref{rex}) and the
corresponding conjugate equality
\begin{equation} \label{xex}
\Delta X_{nc} \delta P= \hbar /2 . 
\end{equation}
The latter exact uncertainty relation 
is proved in a formally equivalent manner; similarly implies the Heisenberg
inequality; requires the variance in position to be infinite for
states with momentum distributions that are discontinuous or which
vanish over a continuous range of momentum values; and implies that
the variance of the nonclassical component of position is strictly
positive.
\newpage
\subsection{Energy bounds}

Eqs. (\ref{pdiff}) and (\ref{rhoeur}) 
immediately yield the lower bound
\begin{equation}\label{enineq}
E\geq (8m)^{-1}\hbar^2(\delta X)^{-2} + \langle V\rangle
\end{equation}
for the average energy $E$ of any state, where from Eq. (\ref{real}) one
has equality for all {\it real wavefunctions}.  Thus
energy bounds may be obtained via bounds on the Fisher length
$\delta X$.

A number of upper and lower bounds for the Fisher length are given by
Dembo et al. \cite{dembo}, and by Romera and Dehesa \cite{romera}, which
hence yield corresponding bounds on energy. 
For example, the ``isoperimetric inequality'' \cite{dembo}
\[
\delta X\leq (2\pi e)^{-1/2} e^S , \]
where $S=-\int dx\,p(x)\ln p(x)$ is the position entropy, 
implies via Eq. (\ref{enineq}) the general
{\it entropic} lower bound
\begin{equation}\label{entbound}
E\geq (4m)^{-1}\pi e\hbar^2 e^{-2S} +\langle V\rangle .
\end{equation}

Eq. (\ref{entbound}) may be exploited to estimate groundstate energies,
by maximising the position entropy for a given value of $\langle
V\rangle$.  Note this gives a lower bound on $E_0$, in contrast to the
usual upper bounds provided by variational methods.  For example, for a
harmonic oscillator with $V(x)=m\omega^2x^2/2$, the entropy is well known
to be maximised for a given value of $\langle x^2\rangle$ by a Gaussian
distribution.  Substituting such a distribution into Eq.
(\ref{entbound}) and minimising with respect to $\langle x^2\rangle$
then yields the estimate $E_0\geq \hbar\omega /2$, where 
the righthand side is in fact the correct groundstate energy 
(because the groundstate probability distribution is indeed 
Gaussian).

As a nontrivial example of Eq. (\ref{entbound}), consider a particle
bouncing in a uniform gravitational field, with $V(x)=mgx$ for $x\geq
0$.  For a fixed value $\langle x\rangle =\lambda$ one finds that the
entropy is maximised by the exponential distribution $p(x)= \lambda^{-1}
\exp (-x/\lambda)$ ($x\geq 0$), yielding the lower bound
\[
E\geq \pi\hbar^2(4me\lambda^2)^{-1} +mg\lambda .\]
Minimizing with respect to $\lambda$ then gives the
estimate
\[
E_0\geq (3/2)[\pi/(2e)]^{1/3}(mg^2\hbar^2)^{1/3}\approx 1.249\, 
(mg^2\hbar^2)^{1/3} , \]
which is comparable to the exact value of
$(mg^2\hbar^2/2)^{1/3}a_0 \approx 1.856\, (mg^2\hbar^2)^{1/3}$ obtained
by solving the Schr\"{o}dinger equation \cite{flugge}, where $a_0$
denotes the first Airy function zero.  
\newpage
\subsection{Phase, angular momentum and photon number}

The decomposition of angular momentum and photon number into classical
and nonclassical components is discussed in detail elsewhere
\cite{eurlong}.  One finds, for example, that the best estimate of
photon number on state $\rho$, which is compatible with a phase
measurement result $\Phi =\phi$, is given by (cf. Eq. (\ref{pcl}))
\[
N_{cl}(\phi ) =\frac{\langle\phi |N\rho +\rho N|\phi\rangle/2}{\langle
\phi |\rho |\phi\rangle} ,
\]
where $N$ is the photon number operator and $|\phi\rangle$ is the
Susskind-Glogower phase state $\sum_n \exp (in\phi )|n\rangle$.  One
also has an additivity property ${\rm Var}N = {\rm Var}N_{cl} + {\rm
Var}N_{nc}$ analogous to Eq. (\ref{pvar}).  A Fisher length $\delta
\Phi$ for the
phase distribution is defined analogously to Eq. (\ref{fl}) (where
integration is restricted to a reference interval of length $2\pi$), and
satisfies a modified form of Cramer-Rao inequality \cite{eurlong}.
 
The corresponding exact uncertainty relations are
\begin{eqnarray} \label{jexact}
\delta \Phi\Delta J_{nc} & = & \hbar/2 ,\\ \label{nexact}
\delta\Phi \Delta N_{nc} & = & 1/2,
\end{eqnarray}
for phase and angular momentum and for phase and photon number
respectively, and are proved exactly as per Eq. (\ref{rex}) above 
for all pure states.  For more general states described by
density operators the righthand sides become lower bounds.
These exact uncertainty relations are far stronger than the corresponding
Heisenberg-type inequalities \cite{hallphase}. 

\section{ENTANGLEMENT AND CORRELATION}

\subsection{Higher dimensions}

Exact uncertainty relations for vector observables are of interest not
only because the world is not one-dimensional, but because some physical
properties, such as entanglement, require more than one dimension for
their very definition.  It is therefore indicated here how
(\ref{ex}) may be generalised to the case of $n$-vectors ${\bf X}$ and
${\bf P}$.  This case has also been briefly considered in Ref. \cite{hall}.
For simplicity only pure states will be considered.

First, one has the vector decomposition 
\begin{equation} \label{vecdecomp}
{\bf P} = {\bf P}_{cl} + {\bf P}_{nc}
\end{equation}
into classical and nonclassical components,
where ${\bf P}_{cl}$ commutes with ${\bf X}$, and
\begin{equation} \label{vecpcl}
{\bf P}_{cl}({\bf x}) = \langle {\bf x}|{\bf P}_{cl}|{\bf x}\rangle =
\frac{\hbar}{2i}\left( \frac{\nabla\psi}{\psi}-\frac{\nabla\psi^*
}{\psi^*}\right) = \hbar\nabla\left[ \arg\psi\right]
\end{equation}
is the best estimate of ${\bf P}$ from measurement value ${\bf X}={\bf
x}$ for state $\psi$ (one may also derive ${\bf P}_{cl}({\bf x})$ from
a continuity equation or a Wigner function approach, as per Sec. II.C).  

In analogy to Eqs. (\ref{pav}) and (\ref{pvar}) one may derive $\langle
{\bf P}\rangle = \langle{\bf P}_{cl}\rangle$ and the generalized linear
independence property
\begin{equation}\label{pcov}
{\rm Cov}({\bf P}) = {\rm Cov}({\bf P}_{cl}) + {\rm Cov}({\bf P}_{nc}) ,
\end{equation}
where the $n\times n$ covariance matrix of $n$-vector ${\bf A}$ is
defined by the matrix coefficients
\begin{equation}\label{cov}
\left[{\rm Cov}({\bf A})\right]_{jk} = \langle A_jA_k\rangle - \langle
A_j\rangle\langle A_k\rangle .
\end{equation}
Note that since the vector components of ${\bf P}$ commute, as do the
vector components of ${\bf P}_{cl}$, then
\[
[P_{nc}^j,P_{nc}^k] = [P^j-P_{cl}^j,P^k-P_{cl}^k] =
(\hbar^2/i)(\partial_j\partial_k -\partial_k\partial_j)\left[ \arg\psi
\right]=0 ,\]
as claimed in Eq. (\ref{commcomp}).

The notion of Fisher length for one dimension generalises to
the matrix inverse
\begin{equation}
\label{fcov}
{\rm FCov}({\bf X}) := \left\{ \int d^nx\, p({\bf x})[\nabla \ln p({\bf
x})]\, [\nabla \ln p({\bf x})]^T\right\}^{-1} ,
\end{equation}
where ${\bf A}^T$ denotes the vector transpose of ${\bf A}$.  For the
case of one dimension this reduces to the square of the Fisher length
$\delta X$, just as the covariance matrix in Eq. (\ref{cov}) reduces to
the square of $\Delta A$.  Moreover, as per the covariance matrix, 
the matrix in Eq. (\ref{fcov}) is
real, symmetric and nonnegative. Finally, the
matrix is the inverse of the ``Fisher information matrix" of statistical
estimation theory \cite{cox}.  For these reasons ${\rm FCov}({\bf X})$
will be referred to as the {\it Fisher covariance matrix} of ${\bf X}$.
One has the generalized Cramer-Rao inequality \cite{cox}
\begin{equation}\label{covcramer}
{\rm Cov}({\bf X}) \geq {\rm FCov}({\bf X}) ,
\end{equation}
with equality for Gaussian distributions.

One may show, via direct calculation of ${\rm Cov}({\bf P}_{cl})$, 
that the generalized exact uncertainty relation
\begin{equation}\label{covexact}
{\rm FCov}({\bf X})\, {\rm Cov}({\bf P}_{nc}) = (\hbar/2)^2I_n 
\end{equation}
holds for all pure states, where $I_n$ denotes the $n\times n$ unit
matrix.  This exact uncertainty relation, being a symmetric matrix
equality, comprises $n(n+1)/2$ independent equalities.  
The corresponding Heisenberg matrix inequality 
\begin{equation}\label{covheis}
{\rm Cov}({\bf X})\, {\rm Cov}({\bf P}) \geq (\hbar/2)^2I_n ,
\end{equation}
follows immediately from Eqs. (\ref{pcov}), (\ref{covcramer}) and
(\ref{covexact}). 

\subsection{Entangled particles}

Consider now the case of two one-dimensional particles, with respective
position and momentum observables $(X^{(1)},P^{(1)})$ and
$(X^{(2)},P^{(2)})$.  Such a system corresponds to $n=2$ above,
and the corresponding nonclassical momentum components associated with
wavefunction $\psi$ follow from Eqs. (\ref{vecdecomp}) and (\ref{vecpcl})
as
\begin{equation} \label{pentang}
P^{(1)}_{nc} = P^{(1)}-\hbar\frac{\partial \arg\psi(x_1,x_2)}{
\partial x_1},\hspace{1cm} 
P^{(2)}_{nc} = P^{(2)}-\hbar\frac{\partial \arg\psi(x_1,x_2)}{
\partial x_2} .
\end{equation}
For entangled states (e.g., a superposition of two product states), 
it follows that the
nonclassical momentum of particle 1 will typically depend on the position
observable of particle 2, and vice versa.  Hence if some
unitary transformation (e.g., a position displacement) is 
performed on the {\it second} particle, then the 
nonclassical momentum of the {\it first} particle is typically 
changed.

The decomposition into classical and nonclassical components is
therefore essentially nonlocal:  
the decomposition of a single-particle observable 
typically depends upon actions performed on another
particle with which the first is entangled.  Conversely, all such decompositions
are invariant under actions performed
on a second {\it un}entangled particle.  
The nonlocality inherent in quantum entanglement is thus 
reflected by the nonlocality
of classical/nonclassical decompositions.

The exact uncertainty relation corresponding to the decomposition of
momentum in Eq. (\ref{pentang}) is given by the matrix equality of Eq.
(\ref{covexact}), with $n=2$.  This leads to three independent
inequalities, two of which may be chosen as 
as generalizations of the exact uncertainty relation
in Eq. (\ref{ex}) for each individual particle.

There are many ways of choosing the third independent inequality. 
However, 
one particular choice provides an interesting connection with the Pearson
correlation coefficient of classical statistics. The latter 
coefficient is defined for two compatible observables $A$ and $B$, 
in terms of the coefficients $C_{jk}$ of the corresponding
covariance matrix ${\rm Cov}(A,B)$, by \cite{cox}
\begin{equation} \label{pearson}
r_P(A,B) := C_{12}/(C_{11}C_{22})^{1/2} , 
\end{equation}
and provides a measure of the degree to which $A$ and $B$
are linearly correlated. It ranges between -1 (a high degree of
linear correlation with negative slope) and +1 (a
high degree of linear correlation with positive
slope).  One may analogously define the ``Fisher'' correlation
coefficient in terms of the coefficients $C^F_{jk}$ of the corresponding
Fisher covariance matrix
${\rm FCov}(A,B)$, by 
\begin{equation}\label{fisher}
r_F(A,B) := C^F_{12}/(C^F_{11}C^F_{22})^{1/2}  .
\end{equation}
This also provides a measure of correlation ranging between -1 and +1,
and is equal to the Pearson correlation coefficient for all Gaussian
distributions.

The third equality may now be
chosen as the simple correlation relation
\begin{equation} \label{corr}
r_P(P^{(1)}_{nc},P^{(2)}_{nc}) + r_F(X^{(1)},X^{(2)}) = 0 , 
\end{equation}
as may be verified by direct calculation from Eq. (\ref{covexact}).
It is seen that the exact uncertainty relation in Eq.
(\ref{covexact}) thus constrains both uncertainty {\it and}
correlation.
 
For example, if the nonclassical momentum components of particles
1 and 2 are positively correlated then the position observables are
negatively correlated, and vice versa.  More generally, the degree of
nonclassical momentum correlation is seen to be
precisely determined by the degree
of position correlation.  Note for {\it unentangled} particles that
Eq. (\ref{corr}) is trivial:  both the Pearson and the Fisher
correlation coefficients vanish identically.  The exact
uncertainty relation in Eq. (\ref{covexact}) thus reduces in this case
to the exact
uncertainty relations (\ref{rex}) for each particle.  

\subsection{EPR correlations}

A nice example is provided by the approximate EPR
state
\[
\psi(x_1,x_2) = K e^{-(x_1-x_2-a)^2/4\sigma^2} 
e^{-(x_1+x_2)^2/4\tau^2} e^{ip_0(x_1+x_2)/(2\hbar)} ,\]
where $K$ is a normalisation constant and $\sigma <<1<<\tau$ in 
suitable units.  One may then
calculate
\begin{eqnarray*}
\langle X^{(1)}-X^{(2)}\rangle & = & a,\hspace{1cm}
{\rm Var}(X^{(1)}-X^{(2)}) = \sigma^2 << 1 ,\\
\langle P^{(1)}+P^{(2)}\rangle & = & p_0,\hspace{1cm}
{\rm Var}(P^{(1)}+P^{(2)}) = \hbar^2/\tau^2 <<1 ,
\end{eqnarray*}
and hence $\psi$ is an approximate eigenstate of the relative position
and the total momentum, i.e., one may write
\begin{equation}\label{approx}
X^{(1)}-X^{(2)}\approx a,\hspace{1cm} P^{(1)}+P^{(2)}\approx p_0 .
\end{equation}
This state is thus an approximate version of the (nonnormalizable) ket
considered by Einstein, Podolsky and Rosen in connection with
the completeness of the quantum theory \cite{epr}.
 
For state $\psi$ one finds from Eq. (\ref{vecpcl}) that the classical
components of momentum are constant, each being equal to $p_0/2$. 
Hence one has ${\rm Cov}{\bf P}_{nc} ={\rm Cov}{\bf P}$ from Eq.
(\ref{pcov}).  Then, since equality holds in Eq. (\ref{covcramer}) for
Gaussian distributions, the exact uncertainty relation corresponding to
$\psi$ follows from Eq. (\ref{covexact}) as
\begin{equation} \label{epruncert}
{\rm Cov}(X) {\rm Cov}(P) = (\hbar/2)^2 I_n  .
\end{equation}
Eq. (\ref{corr}) reduces to  
(recalling that $r_P$ and $r_F$ are equivalent for
Gaussian distributions) the correlation relation
\[
r_P({\bf X}) + r_P({\bf P}) = 0 .\]
This latter result 
is consistent with Eq. (\ref{approx}), which implies that 
$X^{(1)}$ and $X^{(2)}$ are highly positively correlated for state
$\psi$ [$r_P({\bf X})\approx 1$], while $P^{(1)}$ and $P^{(2)}$ are
highly negatively correlated [$r_P({\bf P})\approx -1$].

It is of interest to consider the effect of measurements on the
approximate EPR state $\psi$.  First, for a position
measurement on particle 2, with result $X^{(2)}=x$, the state of
particle 1 collapses to the wavefunction obtained 
by substituting $x_2=x$ and renormalising.  It follows that the the
classical momentum component $P_{cl}^{(1)}$ remains equal to $p_0/2$.
Hence the momentum decomposition of particle 1 is not altered by
knowledge of $X^{(2)}$.

Conversely, for a momentum measurement on particle 2 with result
$P^{(2)}=p$, one finds via straightforward calculation of the
appropriate Gaussian integrals that the state of particle 1 collapses
to the wavefunction
\[
\psi(x_1|P^{(2)}=p) =
K'e^{-(x_1+a/2)^2/(\sigma^2+\tau^2)/4}e^{i\tilde{p}x_1/\hbar} ,\]
where $K'$ is a normalisation constant and 
\[
\tilde{p} = \frac{\sigma^2p+\tau^2(p_0-p)}{\sigma^2+\tau^2} 
\approx p_0-p.\]
It follows that the classical momentum component $P^{(1)}_{cl}$ 
is {\it not} invariant
under a measurement of $P^{(2)}$, changing from $p_0/2$ to
$\tilde{p}$.  Hence there is a ``nonlocal'' effect on the
classical/nonclassical decomposition of momentum for particle 1, brought
about by a measurement of $P^{(2)}$. Thus 
the strong correlation between $P^{(1)}$ and $P^{(2)}$ for
state $\psi$ in Eq. (\ref{approx}) can be considered nonlocal in nature.
Similar results hold with respect to the position correlation. 

\section{OTHER OBSERVABLES}

Exact uncertainty relations can be formally extended in a very
general way to arbitrary pairs of Hermitian observables.  Unfortunately,
the physical significance of such an extension is not entirely
clear, as will be seen below.  However, for the case of a complete set of
mutually complementary observables on a finite Hilbert space, such as
the Pauli spin matrices, it will be
shown that results in the literature provide a very satisfactory
form of exact uncertainty relation.

First, consider the case of {\it any} two observables $A$ and $B$
represented by Hermitian operators, and for state $\rho$ define
\begin{equation}\label{bcl}
B^A_{cl} := \sum_a |a\rangle\langle a|\frac{\langle a|B\rho +\rho
B|a\rangle/2}{\langle a|\rho |a\rangle} .
\end{equation}
Here $|a\rangle$ denotes the eigenket of $A$ with eigenvalue $a$, and
the summation is replaced by integration for continuous ranges of
eigenvalues.

Clearly the above expression generalises Eqs. (\ref{pcl}) and
(\ref{pclop}), and indeed $B^A_{cl}$ may be interpreted as providing the
best estimate of $B$ compatible with measurement of $A$ on state $\rho$.
Note that $A^A_{cl}=A$, i.e., $A$ is its own best estimate.
One may further define $B^A_{nc}$ via the decomposition
\[ B = B^A_{cl} + B^A_{nc}, \]
and obtain the relations
\[
\langle B\rangle = \langle B^A_{cl}\rangle, \hspace{1cm} {\rm Var}B =
{\rm Var}B^A_{cl}+{\rm Var}B^A_{nc} \]
for state $\rho$, in analogy to Eqs. (\ref{pav}) and (\ref{pvar}).

If one is then prepared to define the quantity $\delta_B A$ by
\[
(\delta_B A)^{-2} = \sum_a \frac{\langle a|(i/\hbar)[B,\rho]|a\rangle^2
}{\langle a|\rho |a\rangle} , \]
in analogy to Eq. (\ref{fisho}), then precisely as per the derivation of
Eq. (\ref{rhoeur}) one may show that
\begin{equation} \label{aexact}
(\delta_B A)\,\Delta B^A_{nc} \geq \hbar/2,
\end{equation}
with equality for all pure states.

Thus there is a very straightforward generalisation of Eq. (\ref{ex}) to
arbitrary pairs of observables.  A difficulty is, however, to provide
a meaningful statistical interpretation of $\delta_BA$.  Note in
particular that, unlike the Fisher length $\delta X$, this quantity is
not a functional of the probability distribution $\langle a|\rho
|a\rangle$ in general.  Possibly, noting the commutator which appears in
the definition of $\delta_BA$, one can interpret this quantity as a
measure of the degree to which a measurement of $A$ can distinguish
between $B$-generated translations of state
$\rho$, i.e., between unitary transformations of the form $e^{ixB/\hbar}\rho
e^{-ixB/\hbar}$ \cite{helhol}.  Here such an attempt will not be made,
although it is noted that the case of arbitrary quadrature observables
of a single-mode field should provide a simple test.  

Finally, it is pointed out that a rather different type of
exact uncertainty relation exists for a set of $n+1$ mutually
complementary observables $A_1, A_2, \dots ,A_{n+1}$ on an
$n$-dimensional Hilbert space.  Such sets are defined by the property
that the distribution of any member is uniform for an eigenstate of any
other member, and are known to exist when $n$ is a power of a prime
number \cite{wootters}.  
As an example one may choose $n=2$, and take $A_1$, $A_2$
and $A_3$ to be the Pauli spin matrices.

Let $L$ denote the collision length
of probability distribution $\{p_1, p_2, \dots p_n\}$, defined by \cite{hell}
\[
L := 1/ \sum_j (p_j)^2 . \]
Note that $L$ is equal to 1 for a distribution concentrated on a single
outcome, and is equal to $n$ for a distribution spread uniformly over
all $n$ possible outcomes. It hence provides a direct measure of the spread
of the distribution over the space of outcomes \cite{hell}.

One may show that \cite{ivan}
\begin{equation} \label{mut}
\sum_i 1/L_i = 1+{\rm tr}[\rho^2] \leq 2 ,
\end{equation}
where $L_i$ denotes the collision length of observable $A_i$ for state
$\rho$.  
This reduces to strict equality for all pure states, and 
thus provides an exact uncertainty relation for the
collision lengths of any set of $n+1$ mutually complementary
observables.  For example, if $L_j=1$ for some
observable $A_j$ (minimal uncertainty), then $L_i=n$
for all $i\neq j$ (maximal uncertainty).
Ivanovic has shown that Eq. (\ref{mut}) can be
used to derive an entropic uncertainty relation for the $A_i$
\cite{ivan}, while Brukner and Zeilinger have interpreted Eq.
(\ref{mut}) as an additivity property of a particular 
``information'' measure \cite{zeil}.
\newpage
\section{CONCLUSIONS} 

The existence of exact uncertainty relations connecting the statistics of
complementary observables greatly strengthens the usual statement of the
uncertainty principle:
the lack of knowledge about an observable, for any
wavefunction, is {\it precisely} determined by the lack of knowledge about
the conjugate observable.  The measures of lack of knowledge must of
course be chosen appropriately (as the nonclassical fluctuation strength
and the Fisher length).  What is remarkable is that such measures
can be chosen at all.

The decomposition of the momentum observable into classical and
nonclassical components has a number of clear
physical consequences.  The classical component characterises that part
of the momentum comeasurable with position, while the nonclassical component
successfully characterises the ``quantum" nature of the momentum
observable $P$ (including the exact uncertainty relation
Eq. (\ref{ex}), the nonclassical part of the kinetic energy, and the
nonlocality inherent in the momentum correlations of entangled
particles).  It has been shown elsewhere that the nonclassical
position and momentum uncertainties characterises the robustness of 
quantum systems with respect to Gaussian noise processes \cite{hall},
and the notion of a nonclassical momentum fluctuation inversely
related to position uncertainty has been successfully 
used as a starting point for
{\it deriving} the Schr\"{o}dinger equation \cite{hallreg}.

The exact uncertainty relations in Eqs. (\ref{ex}), (\ref{jexact}),
(\ref{nexact}) and (\ref{covexact}) are formal consequences of the
Fourier transformations which connect the representations of conjugate
quantum observables.  Hence they may be extended to any domain in which
such transformations have physical significance.  This includes 
the time-frequency domain, discussed elsewhere \cite{eurlong} 
(where the "classical" component of
the frequency is essentially the so-called ``instantaneous frequency"
\cite{fante}), as well as
Fourier optics and image processing.

It would be of interest to determine whether exact uncertainty relations
exist for relativistic systems.  One is hampered in direct attempts 
by difficulties associated with one-particle interpretations of the
Klein-Gordon and Dirac equations.  It would perhaps therefore be more
fruitful to first consider extensions to general field theories.

Finally, note that the definition of the Fisher covariance matrix in Eq.
(\ref{fcov}) suggests an analogous definition of a ``Wigner'' covariance
matrix ${\rm WCov}$, defined via the coefficients of its matrix inverse
\[ [{\rm WCov}^{-1}]_{jk} := \int d^{2n}z\, W^{-1} \frac{\partial
W}{\partial z_j} \frac{\partial W}{\partial z_k} .\]
Here $W$ denotes the Wigner function of the state, and ${\bf z}$ denotes
the phase space vector $({\bf x},{\bf p})$.  It would be of interest to
determine to what degree this matrix is well-defined, and to what extent 
its properties characterise nonclassical features of quantum
states. 

{\bf Acknowledgment}
I thank Marcel Reginatto for constant encouragement and many helpful comments on
the subject matter of this paper.
%\newpage


\begin{thebibliography}{99}

\bibitem[1]{heisbohr} W. Heisenberg, Z. Physik {\bf 43}, 172 (1927); 
N. Bohr, {\it Atomic Physics and Human Knowledge} (Wiley, New York,
1958), pp. 32-66.
\bibitem[2]{hall} M.J.W. Hall, Phys. Rev. A {\bf 62}, 012107 (2000).
\bibitem[3]{hallreg} M.J.W. Hall and M. Reginatto, e-print
quant-ph/0102069.
\bibitem[4]{ivan} I.D. Ivanovic, J. Phys. A {\bf 25}, L363 (1992).
\bibitem[5]{merz} E. Merzbacher, {\it Quantum Mechanics}, 2nd ed.
(Wiley, New York, 1970), Sec. 4.1.
\bibitem[6]{wig} M. Hillery, R.F. O'Connell, M.O. Scully, and E.P.
Wigner, Phys. Rep. {\bf 106}, 121 (1984).
\bibitem[7]{takbrown} M.R. Brown, e-print quant-ph/9703007; T.
Takabayasi, Prog. Theor. Phys. {\bf 11}, 341 (1954).
\bibitem[8]{eurlong} M.J.W. Hall, e-print quant-ph/0103072.
\bibitem[9]{hell} E. Heller, Phys. Rev. A {\bf 35}, 1360 (1987).
\bibitem[10]{hallvol} M.J.W. Hall, Phys. Rev. A {\bf 59}, 2602 (1999).
\bibitem[11]{cox} M. J. Schervish, {\it Theory of Statistics}
(Springer-Verlag, New York, 1995), pp. 111, 301-306, 613. 
\bibitem[12]{fish} R.A. Fisher, Proc. Cambridge Philos. Soc. {\bf
22}, 700 (1925).
\bibitem[13]{stam} A.J. Stam, Inf. Control {\bf 2}, 101 (1959).
\bibitem[14]{dembo} A. Dembo, T.M. Cover, and J.A. Thomas, IEEE Trans.
Inf. Theory {\bf 37}, 1501 (1991).
\bibitem[15]{bbm} I. Bialynicki-Birula and J. Mycielski, Commun. Math.
Phys. {\bf 44}, 129 (1975).
\bibitem[16]{footconj} It was conjectured in Ref. \cite{hall} that the
{\it joint} nonclassicality $\Delta X_{nc}\Delta P_{nc}$ is not only
strictly positive, but is bounded below by
$\hbar/2$ for the case of pure states.  This conjecture is, however,
violated by
 the wavefunction
$K\exp[-x^2+i\alpha x^2]$, where $K$ and $\alpha$ are real constants.
\bibitem[17]{romera} E. Romera and J.S. Dehesa, Phys. Rev. A {\bf 50},
256 (1994).
\bibitem[18]{flugge} S. Fl\"{u}gge, {\it Practical Quantum Mechanics
I} (Springer-Verlag, Berlin, 1971), pp. 101-105.
\bibitem[19] {hallphase} D.T. Pegg and S.M. Barnett, Phys. Rev. A {\bf
39}, 1665 (1989); J.H. Shapiro and S.R. Shephard,
Phys. Rev. A {\bf
43}, 3795 (1991); M.J.W. Hall, J. Mod. Opt. {\bf 40}, 809 (1993).
\bibitem[20]{epr} A. Einstein, B. Podolsky, and N. Rosen, Phys. Rev.
{\bf 47}, 777 (1935).
\bibitem[21]{helhol} C.W. Helstrom, {\it Quantum Detection and
Estimation Theory} (Academic Press, New York, 1976); A.S. Holevo, {\it
Probabilistic and Statistical Aspects of Quantum Theory}
(North-Holland, Amsterdam, 1982).
\bibitem[22]{wootters} W.K. Wootters and B.D. Fields, Ann. Phys.
(N.Y.) {\bf 191}, 363 (1989).
\bibitem[23]{zeil} \u{C}. Brukner and A. Zeilinger, Phys. Rev. A {\bf
63}, 022113 (2001).  See also M.J.W. Hall, e-print quant-ph/0007116;
\u{C}. Brukner and A. Zeilinger, e-print quant-ph/0008091.
\bibitem[24]{fante} R.L. Fante, {\it Signal Analysis and Estimation}
(Wiley, New York, 1988), pp. 21, 91.
\end{thebibliography}
\end{document}